\documentclass[twocolumn,prl,twocolumn,showpacs]{revtex4}

\usepackage{graphicx}
\usepackage{dcolumn}

\usepackage{bm}

\begin{document}

\title{\bf Ab-inito study on different phases of ferromagnetic CeMnNi$_{4}$}

\author{P. Murugan,$^1$ Abhishek Kumar Singh,$^1$ G. P. Das,$^2$ 
and Yoshiyuki Kawazoe$^1$ }
\affiliation{
$^1$Institute for Materials Research, 
Tohoku University, Aoba-ku, Sendai 980-8577, Japan \\
$^2$Indian Association for Cultivation of Science, Jadavpur, Kolkata 700032, India }

\date{\today}

\begin{abstract}
Using first-principles density functional calculations, we study the possible phases of CeMnNi$_{4}$ and show  that the ground state is ferromagnetic. We observed the hexagonal phase to be lowest in energy whereas experimentally observed cubic phase lies slightly higher in energy. We optimized the structure in both phases and in all different magnetic states to explore the possibility of the structural and magnetic phase transitions at ground state. We do not find any phase transitions between the magnetic and non-magnetic phases. The calculated structural, magnetic properties of cubic phase are in excellent agreement with experiments. Further, we do not observe half metallic behavior in any of the phases. However, the cubic phase does have fewer density of states for down-spin component giving a possibility of forming half metallic phase artificially, introducing vacancies, and disorder in lattice. 
\end{abstract}

\maketitle

The intermetallic compound, CeNi$_{5}$ crystalizes in hexagonal structure (space group 191). It is a Stoner enhanced paramagnet, which is characterized by the influence of spin fluctuations on the transport properties \cite{reiffers}. Replacements of Ni atoms partially by transition metal elements, like Fe, Mn, Cu in CeNi$_{5}$ expands the lattice and produces the significant changes in the magnetic properties \cite{pourarian, marcano}. Even, by substituting with a non-transition metal elements, like Ga in CeNi$_5$ can change the magnetic properties \cite{tang}. Recently, Singh {\it et al} \cite{singh} observed that cubic CeMnNi$_{4}$ (with lattice parameter of 6.987 {~\AA}, space group 216) is ferromagnetic with a large magnetic moments of 4.95 $\mu_{B}$/Mn atom, which is close to integer moment. Also, this system is found to have a high degree of spin polarization. This makes it potentially important material for spintronics applications, as conduction of electrons in such materials can be manipulated by the application of applied magnetic field. The origin of such large spin polarization in cubic CeMnNi$_{4}$ has not been understood clearly. However, it is possible to understand it by first principles calculations. Moreover, the structural phase transition from hexagonal to cubic was observed in CeMn$_{x}$Ni$_{5-x}$ at higher concentration of Mn in range of $x$ = 0.9 - 2.1 \cite{kalychak}. This indicates the possibility of manifestation of hexagonal structure at $x$ = 1. Hence, we embarked upon detailed first principles investigation of structural and magnetic properties for the hexagonal as well as the cubic phase of CeMnNi$_{4}$. In order to understand the magnetic phase transitions, we carried out calculations on various possible magnetic configurations, viz. ferro-, ferri-, anti-ferro-, as well as non-magnetic phase. Our calculations show that the hexagonal phase is lower in energy compared to cubic phase, while the magnetic and structural properties of cubic phase have excellent matching with the experiment \cite{singh}.

We have used {\it ab initio} projected augmented wave method~\cite{bloechl} and a plane wave basis set within the spin-polarized density functional theory and the generalized gradient approximation (GGA)~\cite{perdew} for the exchange-correlation energy.  The pseudopotentials are taken as implemented in the Vienna Ab Initio Simulation Package (VASP) ~\cite{kresse} . For the $f$-band Ce, the semi-core $s$ and $p$ electrons are also taken as valence. For Mn and Ni, the semi-core $3d$ and $4s$ are considered as valence electrons. For representing the Brillouin zone, we use mesh of 8x8x8 Monkhorst-Pack {\bf k}-point for cubic and 8x8x4 including gamma point for hexagonal phase. Results on convergence test of these phases with respect to k-point sampling shows them to be sufficient within an accuracy of 0.001 eV. The conjugate gradient technique is used to optimize the structures without any symmetry constraints. The structures are considered to be converged when the force on each ion becomes 0.001 eV/{\AA} or less. This high accuracy is also necessary to differentiate between closely lying magnetic phases. Various magnetic phases have been acheived by varying the initial magnetic moments of Mn atoms. We find that the direction of moment in Mn atoms is preserved after relaxations. 

The details of the structural parameter, magnetic moments at equilibrium volume, are given in Table 1. For hexagonal phase we have considered a supercell of 1x1x2 dimensions, having two Mn atom per super-cell this also allows us to perform anti-ferromagnetic calculations. The ionic positions and unit-cell parameters have been relaxed completely. This include optimization of $a$ as well as $c/a$ ratio in hexagonal phase. The optimized structure remains the same with $a$ = 4.93{~\AA} and $c/a$ ratio =0.835.The ground state structure is ferromagnetic as shown in the table. The anti-ferromagnetic phase lies 14 meV/formula unit higher in energy whereas non-magnetic phase lies at 0.70 eV/formula unit, clearly showing preference for magnetic phases in CeMnNi$_{4}$. However, we can not expect a higher T$_{c}$ value. 

\begin{table}
\caption{The structural and magnetic properties of cubic and hexagonal phases of CeMnNi$_{4}$.}
\begin{ruledtabular}
\begin{tabular}{ccc}
CeMnNi$_{4}$ & cubic & hexagonal \\
\hline
space group & F43m & P6/mmm \\
lattice parameter & a=6.99 \AA & a=4.93 {\AA} \\
                            &                      & c/a = 0.835 \\
Total moment  ($\mu_{B}$/Mn) & 4.85 & 3.70 \\  
I (eV/formula unit) & 1.67 & 0.7 \\
J (meV/formula unit) & 47 & 14 \\                      
\end{tabular}
\end{ruledtabular}
\end{table}

Cubic phase has an optimized unit-cell of a= 6.99{~\AA} and is ferromagnetic, like the hexagonal phase, with a magnetic moment of 4.85 $\mu_{B}$/Mn. These are in good agreement with experimental value. Unlike hexagonal phase, the anti-ferromagnetic phase lies 47 meV/formula unit higher in energy. This value could be translated into T$_c$ of 148 K which is very close to reported experimental value~\cite{singh}. The  ferri-magnetic phase with moment of 2.49 $\mu_{B}$/Mn lies at 34 meV/formula unit, whereas the non magnetic phase lies 1.67 eV/formula unit higher in energy. The Stoner parameter (I) defined as energy difference between non-magnetic and ferro-magnetic phases, for cubic  is much higher than the hexagonal phase (Table 1).  Similar trend has been observed for J-exchange parameter defined as energy difference between ferromagnetic and anti-ferromagnetic phase. These results clearly show that the higher magnetic state is well stabilized in cubic phase. 

 Figure 2 shows the energy/formula unit $vs$ lattice parameter for cubic and hexagonal phases. For cubic phase the minima for non-magnetic case lies at different unit cell parameter than the other magnetic states. This shows that at high pressure and / or temperature the magnetic moment of ferromagnetic phase can be completely quenched to a non-magnetic state. However, for the hexagonal phase, we do not observe such phase transition. 

 Figure 3 shows the total and partial density of states of cubic and hexagonal phases. Both phases are metallic with finite number of states at fermi level. The origin of metallicity in both cases are different. In the case of cubic phase the metallicity is due to finite number of states coming from the Ni atoms only. However, in the case of hexagonal phase Mn as well as Ni atoms are contributing for the metallicity. Ce $f$-orbital are unoccupied ($\approx$1eV above the Fermi level) in both cases, and hence is not expected to affect the results of our LDA calculations. In cubic phase the up-spin states are completely occupied whereas down spin states is occupied with the fewer number of states. This also explain the observed polarization in Andreev reflection experiment~\cite{singh}. The oxidation state of Mn in both cubic and hexagonal phase is most likely +2. The local moments on Ce, Mn and Ni for cubic (hexagonal) phase are -0.23 (-0.44), 3.96 (3.04), and 0.28 (0.26) $\mu_B$, respectively. It shows that the net moment on these systems mainly originates from the Mn atoms, as expected. In the cubic phase there is possibility to convert it into half metallic phase by artificially introducing the defects or changing the concentrations of Mn. In that case there is possibility of cubic phase to lie lower in energy compare to hexagonal phase. There is finite possibility of 100 percent spin polarization if the spin conducts through Ce-Mn channel due to the presence of half metallicity. Moreover the bond distance between Ce-Mn (3.03 {\AA}) is quite comparable to Ce-Ni (2.90 {\AA}) so the chances are more here for this channel to conduct. A detail study on possible ways to convert it into a complete half metallic phase via artificial means are currently going on and details would be published elsewhere.

 We have also performed the magnetic phase transition study on cubic phase by changing the pressure via the lattice parameter (Fig. 4). We find that there is no phase transitions between any magnetic or non-magnetic phases. However, the stoner exchange parameter increases with increasing the lattice parameter (Fig. 4). In the other case the J-exchange parameter decreases with the increasing lattice parameter due to the larger distance between Mn and other atoms, which reduces the hybridization~\cite{anderson}. 

 Further, we calculate the energy band structure of both phases. In the case of cubic phase there are very few states (Fig. 5), which are crossing the fermi level corresponding to up-spin component. As discussed above by introducing some other artificial effects these states could be shifted up (valence band states) and down (conduction band states) to transform this system to a half metallic state. However, in the case of hexagonal phase the situation near the fermi energy is more complicated as there are too many states crossing the fermi energy. So it may be difficult to transform this hexagonal phase into half metallic state even though this phase is energetically favorable. Transforming the phase to cubic would be one of the ways to do so as also the phase transition from hexagonal to cubic phase is observed in CeMn$_{x}$Ni$_{5-x}$ in range of  $x$ =0.9 to 2.9.  
 
 In summary, we study the CeMnNi$_{4}$ in both hexagonal and cubic phases considering various possible magnetic states. Our results are in very good agreement with the recent transport spectroscopy experiments. We find ferromagnetic phase to be lowest in energy for both the phases. The hexagonal phase lies lower in energy compare to cubic phase by 0.24 eV/formula unit. Both of the phases are metallic, though the density of states and band structure shows a higher degree of spin polarization in cubic phase. There is a finite possibility of transforming it to half metallic phase by artificial means. This could be a promising material for future spintronics applications. 

\acknowledgments

The authors would like to thank the staffs of the Center for Computational Materials Science
at IMR for the use of the Hitachi
SR8000/64 supercomputing facilities. We would also like to thank Dr. P. Raychaudhuri and Dr. S.K. Dhar for many helpful discussions and also for providing their experimental results before publication. AKS is grateful for the support of JSPS.

\newpage

\begin{figure}
\includegraphics*[width=80mm,angle=0,origin=c,clip=false]{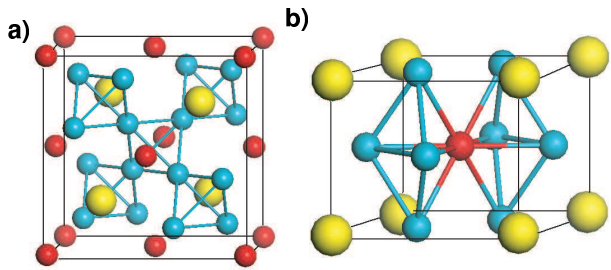}
\vspace{5mm}
\caption{
 a) Cubic and b) hexagonal structure of CeMnNi$_{4}$. Yellow, red, and blue balls represent Ce, Mn, and Ni, respectively.
}
\end{figure}

\begin{figure}
\includegraphics*[width=80mm,angle=0,origin=c,clip=false]{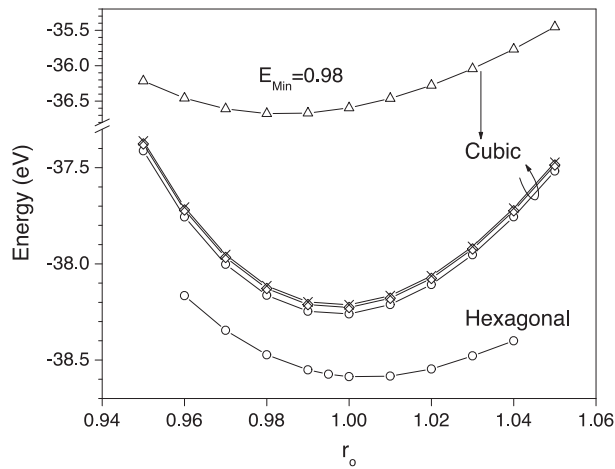}
\vspace{5mm}
\caption{
Plots of total energy/formula unit of CeMnNi$_{4}$ vs normalized lattice parameter in different structural and magnetic phases. The triangle, circle, rhombus and cross corresponds to non-magnetic, ferro, ferri, and anti-ferromagnetic phases, respectively. 
}
\end{figure}

\begin{figure}
\includegraphics*[width=80mm,angle=0,origin=c,clip=false]{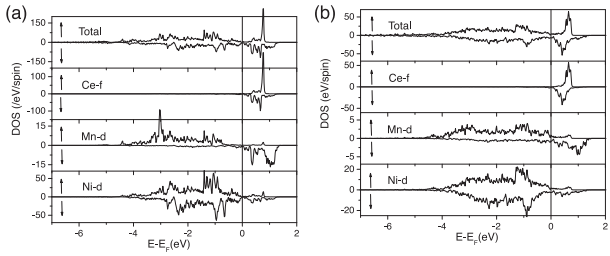}
\vspace{5mm}
\caption{
Total and partial density of states of a) cubic, and b) hexagonal phases are shown. The fermi level are shown by vertical line.The up- and down-spin states are shown by arrow.}
\end{figure}

\begin{figure}
\includegraphics*[width=80mm,angle=0,origin=c,clip=false]{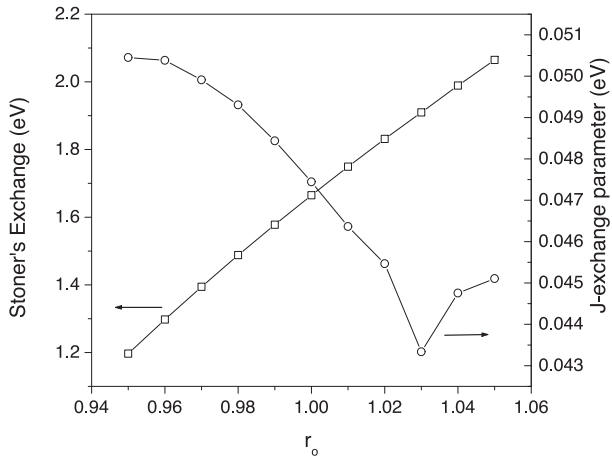}
\vspace{5mm}
\caption{
Plot of Stoner and J-exchange parameters vs normalized lattice parameter for cubic CeMnNi$_{4}$.}
\end{figure}

\begin{figure}
\includegraphics*[width=80mm,angle=0,origin=c,clip=false]{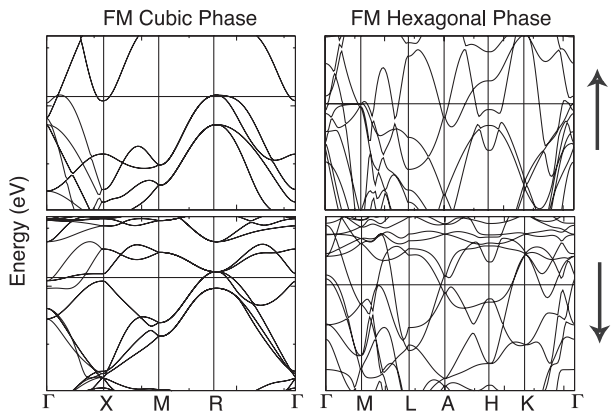}
\vspace{5mm}
\caption{
Band structures of a)cubic and b) hexagonal phases of CeMnNi$_{4}$ for ferromagnetic case. The up- and down-spin states are shown by arrow.}
\end{figure}

\end{document}